%% file: 4p5.tex
\documentclass[
	aps, prl, reprint,
	10pt, notitlepage,
        floats, floatfix,
	amsmath, amssymb, amsfonts, 
	superscriptaddress,
	showpacs, showkeys,
	nofootinbib,
 	longbibliography,
]{revtex4-1}

\def\paper{paper}
\renewcommand{\section}[1]{\paragraph{#1. ---}\phantomsection\addcontentsline{toc}{section}{#1}}

\usepackage{graphicx} 
\usepackage[usenames,dvipsnames]{xcolor}
\usepackage{xspace} 
\usepackage{bm} 
\usepackage{amsmath,amsfonts,amssymb,amsthm}
\usepackage[utf8]{inputenc} 

\xdefinecolor{mylinkcolor}{rgb}{0,0,0.5}
\usepackage[
	bookmarksnumbered, bookmarksopen, bookmarksopenlevel=2,
	breaklinks=true,
	colorlinks=true, filecolor=mylinkcolor, citecolor=mylinkcolor,
	linkcolor=mylinkcolor, urlcolor=mylinkcolor, menucolor=mylinkcolor,
]{hyperref}
\usepackage{verbatim}
\usepackage{mathrsfs}
\usepackage{color}
\usepackage[dvipsnames]{xcolor}
\usepackage{tikz}

\allowdisplaybreaks

\def\vct#1{{\bm{#1}}}

\def\nl{\\ & \quad}
\def\nnl{\nonumber \\ & \quad}

\newcommand{\nnm}{\nonumber}
\newcommand{\doe}{\partial}
\newcommand{\be}{\begin{equation}}
\newcommand{\ee}{\end{equation}}
\newcommand{\bse}{\begin{subequations}}
\newcommand{\ese}{\end{subequations}}

\newcommand{\mr}{\mathrm}

\newcommand{\mc}{\mathcal}

\newcommand{\ms}{\mathsf}

\newcommand{\vinf}{v}

\newcommand{\bpm}{\begin{pmatrix}}
\newcommand{\epm}{\end{pmatrix}}

\newcommand{\AEI}{\affiliation{Max Planck Institute for Gravitational Physics (Albert Einstein Institute), Am M\"uhlenberg 1, Potsdam 14476, Germany}}
\newcommand{\Maryland}{\affiliation{Department of Physics, University of Maryland, College Park, MD 20742, USA}}

\usepackage[normalem]{ulem}

\begin{document}

\title{Gravitational spin-orbit coupling through third-subleading post-Newtonian order: \\ from first-order self-force to arbitrary mass ratios}

\author{Andrea Antonelli}\AEI
\author{Chris Kavanagh}\AEI
\author{Mohammed Khalil}\AEI\Maryland
\author{Jan Steinhoff}\AEI
\author{Justin Vines}\AEI

\begin{abstract}
Exploiting simple yet remarkable properties of relativistic gravitational scattering, we use first-order self-force (linear-in-mass-ratio) results to obtain arbitrary-mass-ratio results for the complete third-subleading post-Newtonian (4.5PN) corrections to the spin-orbit sector of spinning-binary conservative dynamics, for generic (bound or unbound) orbits and spin orientations.
We thereby improve important ingredients of models of gravitational waves from spinning binaries, and we demonstrate the improvement in accuracy by comparing against aligned-spin numerical simulations of binary black holes.
\end{abstract}

\maketitle

\section{Introduction}

The success of gravitational-wave (GW) astronomy in the next decades relies on significantly improved theoretical predictions of GW signals from coalescing binaries of \emph{spinning} compact objects such as black holes (BHs).
A network of GW detectors~\cite{TheLIGOScientific:2014jea,TheVirgo:2014hva} has now observed dozens of signals from binary BHs,
measuring distributions of the BHs' masses and spins and extrinsic properties, 
enabling diverse applications in astro- and fundamental physics~\cite{LIGOScientific:2018mvr,LIGOScientific:2018jsj,LIGOScientific:2019fpa,Abbott:2019yzh}: e.g., discerning binary BH formation channels~\cite{LIGOScientific:2018jsj}, measurement of the Hubble constant~\cite{Abbott:2019yzh}, and tests of general relativity (GR)~\cite{LIGOScientific:2019fpa}.
The search for and parameter estimation of GW signals require accurate predictions, from the inspiral (treated by analytic approximations) to the last orbits and merger of the binary (treated by numerical relativity, NR).
The current accuracy of theoretical predictions, from combined analytic and numerical methods, will likely become insufficient when current detectors reach design sensitivity around 2022~\cite{Purrer:2019jcp}.
More accurate predictions for gravitational waves are thus key to enable the 
physics applications mentioned above.

The primary relevant analytic approximation is the post-Newtonian (PN, weak-field and slow-motion) approximation.
The conservative orbital dynamics is known for nonspinning binaries to the fourth-subleading PN order~\cite{Damour:2014jta,Bernard:2016wrg,Foffa:2019rdf,Foffa:2019yfl,Blumlein:2020pog} (with partial results at the fifth~\cite{Ledvinka:2008tk,Blanchet:2018yvb,Foffa:2019hrb,Blumlein:2019zku,Bini:2019nra} and sixth~\cite{Blumlein:2020znm,Cheung:2020gyp,Bini:2020nsb}), but only to second-subleading order (or next-to-next-to-leading order, N$^2$LO) in the spin-orbit sector~\cite{Hartung:2011te,Levi:2015uxa,Bohe:2012mr}.  The gravitational spin-orbit couplings, linear in the component bodies' spins, are analogous to those in atomic physics.
Recently, the three-loop Feynman integrals at N$^3$LO in the spin-orbit case were calculated~\cite{Levi:2020kvb}, leaving however plenty of tensorial lower-loop integrals as a comparably large computational task.
Innovations that complement these massive algebraic manipulations are thus of great potential value.

In this {\paper}, we follow a line of reasoning which leads to a complete result for the sought-after N$^3$LO-PN spin-orbit dynamics (at 4.5PN order for rapidly spinning binaries), requiring relatively little computational effort by building on a diverse array of previous results.
We extend to the spinning case a novel approach based on special properties of the gauge-invariant scattering-angle function~\cite{Bini:2019nra,Damour:2019lcq,Vines:2018gqi}, which encodes the complete binary dynamics (both bound and unbound).
The weak-field approximation of the scattering angle is strongly constrained by results in the small-mass-ratio approximation\footnote{We define the small-mass-ratio limit as $q=\frac{m_1}{m_2}\ll1$, where $m_{1,2}$ are the masses of the compact objects.}, as treated in the gravitational \emph{self-force} paradigm~\cite{Barack:2018yvs}.
The scattering-angle constraints imply that known first-order (linear-in-mass-ratio) self-force results with spin~\cite{Bini:2016dvs,Kavanagh:2016idg,Kavanagh:2017wot} uniquely fix the full N$^3$LO-PN spin-orbit dynamics for arbitrary mass ratios.
This result completes the 4.5PN conservative dynamics of (rapidly) spinning binaries, together with the NLO cubic-in-spin couplings~\cite{Levi:2019kgk} (see also \cite{Siemonsen:2019dsu}).

As applications, we compute quantities which can be employed to improve waveform models for GW astronomy: the circular-orbit aligned-spin binding energy and the effective gyro-gravitomagnetic ratios.
The former is a crucial ingredient in the construction of faithful models (together with the GW energy flux), for which we quantify the accuracy gain due to the present results by comparing to NR simulations. 
The latter parametrize spin effects in the \texttt{SEOBNR} waveform codes~\cite{Bohe:2016gbl,Babak:2016tgq,Cotesta:2018fcv,Ossokine:2020kjp} used in LIGO-Virgo searches and inference analyses~\cite{LIGOScientific:2018mvr} and in the upcoming \texttt{TEOBResumS} waveform models~\cite{Nagar:2018plt,Nagar:2018zoe}.
The gyro-gravitomagnetic ratios are analogous to the famous ``g-factor'' describing the anomalous magnetic dipole moment of the electron, where contributions at the fifth subleading order were obtained~\cite{Aoyama:2012wj} and lead to spectacular agreement with experiment~\cite{Odom:2006zz}.
Regarding the gravitational analog, experimental constraints on the gyro-gravitomagnetic ratios are so far seemingly out of reach. In fact, only two GW events were observed to contain nonvanishing spin effects with 90\% confidence~\cite{LIGOScientific:2018mvr} (see also Refs.~\cite{Zackay:2019tzo,Huang:2020ysn}). 
However, this will change, e.g., when systems with precessing spins are observed in the future, since the precession of the orbital plane leads to a characteristic modulation of the emitted GWs. 
This may allow improved tests of GR and inference of spins.  Measuring BH spins and their orientations is also important for discriminating binary formation channels~\cite{LIGOScientific:2018jsj}.

We begin by extending the link between weak-field scattering and the self-force approximation~\cite{Vines:2018gqi,Bini:2019nra,Damour:2019lcq} to the spin-orbit sector.
Using existing self-force results, we are then able to uniquely determine the N$^3$LO-PN spin-orbit dynamics, as encoded in the gauge-invariant scattering angle.
We continue by calculating the gyro-gravitomagnetic ratios and circular-orbit aligned-spin binding energy.
We compare to NR simulations to quantify the accuracy improvement and present our conclusions.
$G$ denotes Newton's constant, and $c$ the speed of light.

\section{The mass dependence of the scattering angle}

\begin{figure}
  \centering
  \includegraphics[width=0.9\linewidth]{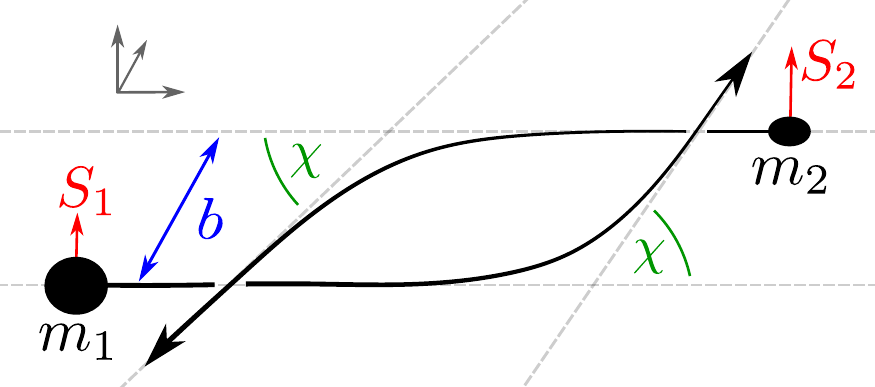}
  \caption{Illustration of aligned-spin scattering BHs.}
  \label{fig:scattering}
\end{figure}

The local-in-time conservative dynamics of a two-massive-body system (without spin or higher multipoles) is fully encoded in the system's gauge-invariant scattering-angle function $\chi(m_1,m_2,v,b)$ ~\cite{Damour:2016gwp,Damour:2017zjx}.  This gives the angle $\chi$ by which both bodies are deflected in the center-of-mass frame, as a function of the masses $m_\ms a$ ($\ms a=1,2$), the asymptotic relative velocity $v$, and the impact parameter $b$.  Based on the structure of iterative solutions in the weak-field (post-Minkowskian) approximation, it has been argued in Sec.~II of Ref.~\cite{Damour:2019lcq} that this function exhibits the following simple dependence on the masses (at fixed $v$ and $b$), through the total mass $M=m_1+m_2$ and the symmetric mass ratio $\nu=m_1m_2/M^2$,
\bse\label{mastereq}
\begin{alignat}{3}\label{chiform}
\frac{\chi}{\Gamma} &= \frac{GM}{b}X_{G^1}^{\nu^0}(v)+\Big(\frac{GM}{b}\Big)^2X_{G^2}^{\nu^0}(v)
\\\nnm
&\quad +\Big(\frac{GM}{b}\Big)^3\Big[X_{G^3}^{\nu^0}(v)+\nu X_{G^3}^{\nu^1}(v)\Big]
\\\nnm
&\quad +\Big(\frac{GM}{b}\Big)^4\Big[X_{G^4}^{\nu^0}(v)+\nu X_{G^4}^{\nu^1}(v)\Big]+\mc O\Big(\frac{GM}{b}\Big)^5,
\end{alignat}
where $\Gamma=E/Mc^2$, with $E^2=(m_1^2+m_2^2+2m_1m_2\gamma)c^4$ being the squared total energy, and $\gamma=(1-v^2/c^2)^{-1/2}$ the asymptotic relative Lorentz factor.  The remarkable fact to be noted here is that the $\mc O(\frac{GM}{b})^{1,2}$ terms are independent of $\nu$, while the $\mc O(\frac{GM}{b})^{3,4}$ terms depend linearly on $\nu$.

As will be argued in detail in future work,${}^{\ref{foot:deriv}}$ this result generalizes straightforwardly to the case of spinning bodies in the aligned-spin configuration, i.e., spins pointing in the direction of the orbital angular momentum (as shown in Fig.~\ref{fig:scattering}).  The aligned-spin dynamics is fully described by the aligned-spin scattering-angle function $\chi(m_\ms a,S_\ms a,v,b)$~\cite{Vines:2018gqi}. Here, $S_\ms a=m_\ms a ca_\ms a$ are the signed spin magnitudes, positive if aligned as in Fig.~\ref{fig:scattering}, negative if anti-aligned.  At the spin-orbit (linear-in-spin) level, the form of Eq.~(\ref{chiform}) holds, with the $X$ functions acquiring additional (linear) dependence on the spins \emph{only} through the dimensionless ratios $a_\ms a/b={S_\ms a}/{m_\ms acb}$, as follows:\footnote{\label{foot:deriv}Note that our Eq.~(\ref{mastereq}) is equivalent to Eqs.~(2.14) and (2.15) of Ref.~\cite{Damour:2019lcq}, but with all the functions $\ms Q^{n\mr{PM}}_{\cdots}(\gamma)$ on the right-hand side of (2.15) replaced by functions $\ms Q^{n\mr{PM}}_{\cdots}(\gamma,a_1/b,a_2/b)$ which are linear in $a_1/b$ and $a_2/b$, and with the additional constraints imposed by symmetry under $(m_1,a_1)\leftrightarrow(m_2,a_2)$.  The arguments leading to this result are very much analogous to those for the spinless case as given in Ref.~\cite{Damour:2019lcq} --- using the structure of the PM expansion, Poincar\'e symmetry, dimensional analysis, etc.\ --- with the given mass dependence holding at fixed ``geometric quantities,'' except that these are now $v,b,a_1,a_2$ instead of just $v$ (or $\gamma$) and $b$.  The rescaled spins $a_\ms a=S_\ms a/m_\ms a c$ and the ``covariant'' (Tulczyjew-Dixon) worldlines (separated by the ``covariant'' impact parameter $b$) are identified as the appropriate geometrical (mass-independent) quantities, because it is in terms of these variables that the first-order metric perturbation is linear in the masses.}
\begin{alignat}{3}\label{replaceX}
X_{G^{n}}^{\nu^m} &\to X_{G^{n}}^{\nu^m}(v)+\frac{a_+}{b}X_{G^na_+}^{\nu^m}(v)+\delta\frac{a_-}{b}X_{G^na_-}^{\nu^m}(v),
\end{alignat}
\ese
where $a_\pm = a_2\pm a_1$ and $\delta=(m_2-m_1)/M$, with the special constraints $X_{G^1a_-}^{\nu^0}=0=X_{G^3a_-}^{\nu^1}$; cf.\ Eq.~(4.32) of Ref.~\cite{Vines:2018gqi}, where this is seen to hold through N$^2$LO in the PN expansion.  It is crucial to note that the impact parameter $b$ in Eq.~\eqref{mastereq}, is the (``covariant'') one orthogonally separating the asymptotic worldlines defined by the Tulczyjew-Dixon condition~\cite{Dixon:1979,Tulczyjew:1959} for each spinning body~\cite{Vines:2018gqi,Vines:2017hyw}.

Now, the fourth order in $G M / b$ encodes the complete spin-orbit dynamics at N$^3$LO in the PN expansion, and according to Eq.~\eqref{mastereq} only terms up to linear order in the mass ratio $\nu$ appear on the right-hand side (noting $\delta\to\pm1$ as $\nu\to0$)---that is, first-order self-force (linear-in-$\nu$) results can be employed to fix the functions $X_{G^n\cdots}^{\nu^m}(v)$ for $n \leq 4$.

\section{Scattering angle, Hamiltonian, and binding energy}

We now connect the scattering angle to an ansatz for a local-in-time binary Hamiltonian including spin-orbit interactions.
If nonlocal-in-time (tail) effects are present, this step requires extra care~\cite{Bini:2019nra}, but this is not the case at the N$^3$LO-PN spin-orbit level.
Crucially, the Hamiltonian describes the dynamics for both unbound (scattering) and bound orbits.
The latter are not only most relevant for GW astronomy, but are also where the vast majority of self-force results are available.
Hence, a gauge-dependent Hamiltonian allows us to connect the scattering angle \eqref{mastereq} with known self-force results.

Let us parametrize our binary Hamiltonian $H(\vct r,\vct p,\vct S_1,\vct S_2)$ in the effective-one-body (EOB)~\cite{Buonanno:1998gg} form,
\begin{equation}\label{HEOB}
  H = M c^2 \sqrt{1 + 2 \nu \left( \frac{H_\text{eff}}{\mu c^2} - 1 \right)} ,
\end{equation}
where $H_\text{eff}(\vct r,\vct p,\vct S_1,\vct S_2)$ is the effective Hamiltonian and $\mu = M \nu$ is the reduced mass, with canonical Poisson brackets $\{r^i,p_i\}=\delta^i_j$, $\{S_\ms a^i,S_\ms a^j\}=\epsilon^{ij}{}_kS_\ms a^k$, and all others vanishing.  At the spin-orbit level, to linear order in the spins, parity invariance implies that $H$ can depend on the spins only through the scalars $\vct L\cdot \vct S_\ms a$, where $\vct L=\vct r\times\vct p$ is the canonical orbital angular momentum.  Thus, a generic Hamiltonian ansatz is of the form
\begin{equation}
  H_\text{eff} = H_\text{eff}^\text{ns} + \frac{1}{c^2 r^3} \vct{L} \cdot (g_S \vct{S} + g_{S^*} \vct{S}^*), \label{Heff}
\end{equation}
where $H_\text{eff}^\text{ns}(\vct r,\vct p)$ is the nonspinning Hamiltonian. We use the conventional spin combinations $\vct{S} = \vct{S}_1 + \vct{S}_2$, $\vct{S}^* = \frac{m_2}{m_1} \vct{S}_1 + \frac{m_1}{m_2} \vct{S}_2$, while $g_S(\vct r,\vct p)$ and $g_{S^*}(\vct r,\vct p)$ are the effective gyro-gravitomagnetic ratios.  In specializing to the case of aligned spins, in which $\vct S_\ms a=S_\ms a\hat{\vct L}$ are (anti)parallel to $\vct L=L\hat{\vct L}$ ($L=|\vct L|$), the motion is confined to the plane orthogonal to the angular momenta, and Eq.~\eqref{Heff} simplifies to
\begin{equation}
  H_\text{eff} =  H_\text{eff}^\text{ns} + \frac{1}{c^2 r^3} L  (g_S S + g_{S^*} S^*), \quad \text{(aligned)} \label{Heffaligned}
\end{equation}
where, crucially, $g_S$ and $g_{S^*}$ are unmodified by this specialization (as they are independent of the spins).  The aligned-spin Hamiltonian is therefore sufficient to reconstruct the generic-spin Hamiltonian, up to the spin-orbit level.  We can adopt polar coordinates $(r,\varphi)$ in the orbital plane, with canonically conjugate momenta $(p_r,L)$, and the Hamiltonian is independent of $\varphi$ due to rotation invariance.  Then $H^\text{ns}_\text{eff}$, $g_S$, and $g_{S^*}$ are each functions of $(r,p_r,L)$.   We take $H_\text{eff}^\text{ns}$ to be given to 4PN order by Eqs.~(5.1) and (8.1) in Ref.~\cite{Damour:2015isa}.  Considering the freedom under canonical transformations, it can be shown that there exists a gauge in which $g_S$ and $g_{S^*}$ are independent of $L$ \cite{Damour:2008qf,Nagar:2011fx,Barausse:2011ys}; we adopt this choice and parametrize our spin-orbit Hamiltonian with the undetermined gyro-gravitomagnetic ratios $g_S(r, p_r)$ and $g_{S^*}(r, p_r)$.
Each term in a PN-expanded ansatz for $g_S$ and $g_{S^*}$ carries a certain power in $c$, from which the PN order can be read off; we include terms up to $c^{-6}$ here.
($c^{-2}$ corresponds to one PN order and $c \rightarrow \infty$ to the Newtonian limit.)

To ascribe physical significance to the spin-orbit Hamiltonian, we point to
the striking similarity with the electromagnetic spin-orbit interactions in atomic physics, which makes $g_S$ and $g_{S^*}$ analogous to the ``g-factor'' of the electron (except that $g_S$ and $g_{S^*}$ depend on dynamical variables). 
This is no accident, since the gravito-magnetic field generated, e.g., by a rotating mass, can be interpreted to exert a Lorentz-like force.
The relativistically preferred geometrical interpretation is that gravito-magnetic fields are dragging inertial/free-falling reference frames, as impressively demonstrated by the Gravity Probe B satellite experiment~\cite{Everitt:2011hp}.

We constrain the ansatz for the Hamiltonian by requiring that it reproduces (i) the mass dependence of the scattering angle \eqref{mastereq}, (ii) the $\nu\to0$ limit of the scattering angle, for a spinning test particle in a Kerr background, as obtained, e.g., by integrating Eq.~(65) of Ref.~\cite{Bini:2017pee}, and (iii) certain gauge-invariant self-force observables, namely, the Detweiler-Barack-Sago redshift~\cite{Detweiler:2008ft,Barack:2011ed,Bini:2016dvs,Hopper:2015icj,Bini:2016qtx,Kavanagh:2015lva,Kavanagh:2016idg,Bini:2015mza} and the spin-precession frequency~\cite{Dolan:2013roa,Akcay:2016dku,Akcay:2017azq,Bini:2014ica,Bini:2015mza,Kavanagh:2017wot,Bini:2018ylh,Bini:2018aps} for bound eccentric aligned-spin orbits, to linear order in the mass ratio.
The scattering angle $\chi$ is obtained from the Hamiltonian \eqref{HEOB} via Eq.~(4.10) of Ref.~\cite{Vines:2018gqi}, with the translation from the total energy $E=H$ and canonical orbital angular momentum $L$ to the asymptotic relative velocity $v$ and ``covariant'' impact parameter $b$ accomplished by Eqs.~(4.13) and (4.17) of Ref.~\cite{Vines:2018gqi}.
The redshifts $z_\ms a$ and spin-precession frequencies $\Omega_\ms a$  ($\ms a=1,2$) are given by
\begin{equation}\label{1law}
  z_\ms a =\bigg\langle \frac{\partial H}{\partial m_\ms a} \bigg\rangle, \qquad
  \Omega_\ms a =\bigg\langle \frac{\partial H}{\partial S_\ms a}\bigg\rangle,
\end{equation}
where $\langle\cdots\rangle$ denotes an average over one period of the radial motion, following from a first law of binary mechanics for eccentric aligned-spin orbits~\cite{LeTiec:2011ab,Blanchet:2012at,Tiec:2015cxa,Fujita:2016igj}.  The procedure for expressing these quantities, in the small-mass-ratio limit, in terms of variables used in self-force calculations is detailed in Ref.~\cite{Bini:2019lcd}.  In this process, to reach the N$^3$LO-PN accuracy in the spin-orbit sector, it is necessary to include the nonspinning 4PN part of the Hamiltonian, including the nonlocal tail part~\cite{Damour:2014jta}, given as an expansion in the orbital eccentricity as in Ref.~\cite{Damour:2015isa}.  After lengthy calculation, working consistently in the small-mass-ratio and PN approximations, we obtain, from our Hamiltonian ansatz, expressions for the redshift $z_1$ and precession frequency $\Omega_1$ of the smaller body, which can be directly compared with the self-force results in Eq.~(4.1) of Ref.~\cite{Kavanagh:2016idg}, Eq.~(23) of Ref.~\cite{Bini:2016dvs} and Eq.~(20) of Ref.~\cite{Bini:2019lcd} for the redshift, and Eq.~(3.33) of Ref.~\cite{Kavanagh:2017wot} for the precession frequency.
The resultant constraints uniquely fix $g_S(r, p_r)$ and $g_{S^*}(r, p_r)$ at N$^3$LO, via an overdetermined system of equations. 

From the Hamiltonian, we can finally calculate the \emph{aligned-spin} circular-orbit binding energy $E_b = H-Mc^2$ as a function of the circular-orbit frequency $\omega=d\varphi/dt=\doe H/\doe L$.
This is a gauge-invariant relation that can be compared to NR.
We decompose $E_b$ into nonspinning and spin-orbit (SO) parts, and further into PN orders, as in
\begin{equation}\label{bindSO}
  E_b^\text{SO} = E_{b,\text{LO}}^\text{SO} + E_{b,\text{NLO}}^\text{SO} + E_{b,\text{N$^2$LO}}^\text{SO} + E_{b,\text{N$^3$LO}}^\text{SO} + \dots . \\
\end{equation}
We can decompose the $g_S$, $g_{S^*}$, and $\chi_\text{SO}$ results from the previous discussion in the same way.
The N$^3$LO pieces of all these quantities are the main results of this {\paper}:
\begin{widetext}
\begin{align}
  \label{chiNNNLOSO}
    \frac{\chi_\text{SO}^\text{N$^3$LO}}{\Gamma}&=
    \frac{\vinf}{c\, b} \Big(a_+ \quad \delta a_-\Big)\Bigg\{ \bigg[\frac{1}{4}\bpm 177 \nu \\ 0 \epm \frac{\vinf^6}{c^6}\bigg] \left(\frac{GM}{\vinf^2b}\right)^3
    + \pi \Bigg[ \frac{3}{4} \bpm -91 + 13\nu \\ -21 + \nu \epm \frac{\vinf^2}{c^2}
    -\frac{1}{8}\bpm 1365 - 777\nu \\ 315 - 45\nu \epm\frac{\vinf^4}{c^4} \nonumber\nl
    \qquad \qquad \qquad \quad -\frac{1}{32} \bpm 1365 - \left(\frac{23717}{3} - \frac{733\pi^2}{8}\right)\nu \\ 315 - \left(\frac{257}{3} + \frac{251\pi^2}{8}\right)\nu \epm\frac{\vinf^6}{c^6} \Bigg] \left(\frac{GM}{\vinf^2b}\right)^4 \Bigg\},\\
    c^6 g_S^\text{N$^3$LO} &= \frac{\nu}{1152} \left(-80399+1446\pi^2+13644\nu-63\nu^2\right)\frac{(GM)^3}{r^3} + \frac{3\nu}{64} \left(-1761+2076\nu+23\nu^2\right) \frac{p_r^2}{\mu^2} \frac{(GM)^2}{r^2} \nnl
      + \frac{\nu}{128} \left(781+3324\nu-771\nu^2\right) \frac{p_r^4}{\mu^4} \frac{GM}{r} + \frac{7\nu}{128} \left(1-36\nu-95\nu^2\right) \frac{p_r^6}{\mu^6} ,\\
    c^6 g_{S^*}^\text{N$^3$LO} &= - \frac{1}{384} \left[1215+2(7627-246\pi^2)\nu-4266\nu^2+36\nu^3 \right] \frac{(GM)^3}{r^3} - \frac{3}{64} \left(15+558\nu-1574\nu^2-36\nu^3\right) \frac{p_r^2}{\mu^2} \frac{(GM)^2}{r^2} \nnl
      + \frac{1}{128} \left(-1105-106\nu+702\nu^2-972\nu^3\right) \frac{p_r^4}{\mu^4} \frac{GM}{r} 
      - \frac{7}{128} \left(45+50\nu+66\nu^2+60\nu^3\right) \frac{p_r^6}{\mu^6} ,\\
    E_{b,\text{N$^3$LO}}^\text{SO} &= - \frac{\nu c^3}{G M} \frac{v_\omega^{11}}{c^{11}} \bigg[ S \bigg( 45 - \frac{19679+174\pi^2}{144} \nu + \frac{1979}{36} \nu^2 + \frac{265}{3888} \nu^3 \bigg)
    + \frac{S^*}{8} \bigg( \frac{135}{2} - 565 \nu + \frac{1109}{3} \nu^2 + \frac{50}{81} \nu^3 \bigg) \bigg] , \label{EbNNNLO}
\end{align}
\end{widetext}
where $v_\omega = (GM\omega)^{1/3}=x^{1/2}c$. 
One needs to add our Eq.~\eqref{chiNNNLOSO} to Eq.~(4.32b) in Ref.~\cite{Vines:2018gqi} to obtain the complete spin-orbit scattering-angle contribution through N$^3$LO-PN and through $\mc O(\frac{GM}{b})^4$.
The lower-order corrections to $E_b^\text{SO}$ can be found in Eq.~(5.4) of Ref.~\cite{Levi:2015uxa}, and the lower-order gyro-gravitomagnetic ratios in Eqs.~(55) and (56) of Ref.~\cite{Nagar:2011fx} (see also Ref.~\cite{Damour:2008qf,Barausse:2011ys}). Through the results for $g_S$ and $g_{S^*}$ presented above, one can straightforwardly improve the \texttt{SEOBNR} waveform models~\cite{Bohe:2016gbl,Babak:2016tgq,Cotesta:2018fcv,Ossokine:2020kjp} used in contemporary gravitational-wave data analysis~\cite{LIGOScientific:2018mvr}.
Likewise, one can use them to improve the upcoming \texttt{TEOBResumS} waveform models~\cite{Nagar:2018plt,Nagar:2018zoe}.
The other main waveform model used by LIGO-Virgo data analysis~\cite{LIGOScientific:2018mvr} is the \texttt{IMRPhenom} family~\cite{Hannam:2013oca,Husa:2015iqa,Khan:2015jqa,Khan:2019kot,Garcia-Quiros:2020qpx,Pratten:2020fqn,Pratten:2020ceb}, which can also be improved using our results, though less directly.

\section{Comparison to NR}
\begin{figure}
\centering
\includegraphics[width=\linewidth]{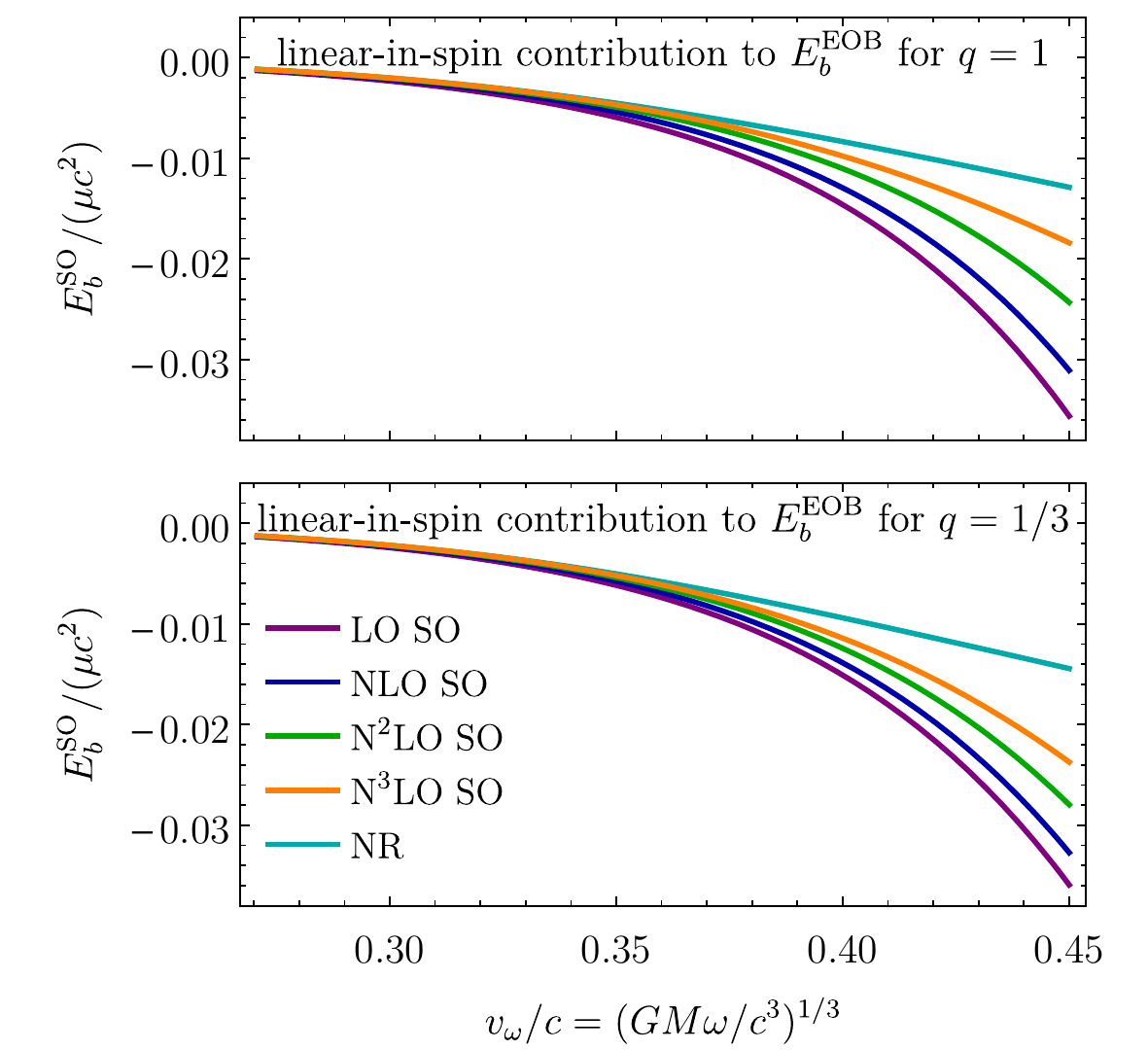}
\caption{
  Comparison of the gauge-invariant relation between the circular-orbit aligned-spin spin-orbit binding energy $E_b$ and $v_\omega$.
  The figure shows results obtained numerically from the (PN-resummed) EOB Hamiltonian \eqref{HEOB} and NR results from Refs.~\cite{Ossokine:2017dge,SXS}.
  The linear-in-spin contribution is isolated using Eq.~\eqref{EbSO} with spin magnitudes $\hat{a} = 0.6$ and for mass ratios $q=1$ and $1/3$.}
\label{fig:bindingenergy}
\end{figure}

We now quantify the improvement in accuracy from the new N$^3$LO spin-orbit correction. The circular-orbit aligned-spin binding energy is a particularly good diagnostic for this, since it encapsulates the conservative dynamics of analytical models, and can be obtained from accurate NR simulations~\cite{Damour:2011fu,Nagar:2015xqa}.
Of particular interest for us is the possibility to (approximately) isolate the linear-in-spin (spin-orbit) contribution by combining the binding energy for two configurations with spins parallel and anti-parallel to the direction of the angular momentum  as follows \cite{Dietrich:2016lyp,Ossokine:2017dge}
\begin{equation}
\label{EbSO}
E_b^\text{SO}(\nu,\hat{a},\hat{a}) = \frac{1}{2} \left[ E_b(\nu,\hat{a},\hat{a}) - E_b(\nu,-\hat{a},-\hat{a})\right],
\end{equation}
with dimensionless spin $\hat{a} = \hat{a}_{\ms a} \equiv c S_{\ms a} / (G m_{\ms a}^2)$.
The result, based on recent NR simulations~\cite{SXS,Ossokine:2017dge}, is shown in Fig.~\ref{fig:bindingenergy}.
The figure also shows the spin-orbit binding energy extracted numerically from the EOB Hamiltonian~\eqref{HEOB}, combining two binding energies for different spin directions in the same way as in the NR case.
The N$^3$LO spin-orbit result shows a clear advantage over the N$^2$LO one, \and that improvement is more pronounced for equal masses than for slightly unequal masses.
[The N$^3$LO PN binding energy (\ref{EbNNNLO}) is very similar to the EOB one for the shown mass ratios.]
This indicates that an inclusion of the N$^3$LO into existing waveform models may lead to improvements even in the strong-field regime, 
otherwise only accessible by computationally-expensive NR simulations.
Recall that gravitational waves are observed from low frequencies (where approximation methods are applicable) to high frequencies (where PN theory is expected to break down).

\section{Conclusions}
Currently-operating (second-generation) gravitational-wave detectors require accuracy improvements for GW predictions by the time they reach design sensitivity around 2022, which become even more stringent for future upgrades and the upcoming third generation of detectors~\cite{Purrer:2019jcp}.
The detector upgrades~\cite{LIGOaplus} in the coming years and a concurrent growing network of observatories~\cite{Aso:2013eba,LIGOIndia} also imply an increased number of detections~\cite{Aasi:2013wya}, making it overall more likely to observe binaries oriented ``edge on'' instead of ``face on,'' which allows measuring precession and extracting spin values with higher accuracy.
The accurate modeling of GW modulations caused by precession, and also the phase accuracy in the aligned-spin case and the contingent improvement in the estimation of spin parameters, motivate us to push predictions for gravitational spin effects to higher orders.

For this purpose, we extended to spin-orbit couplings a link between the weak-field and small-mass-ratio approximations, via the scattering-angle function, as proposed in the nonspinning case in Ref.~\cite{Bini:2019nra,Damour:2019lcq} (see also Ref.~\cite{Vines:2018gqi}).  We employed existing self-force results~\cite{Bini:2016dvs,Kavanagh:2016idg,Kavanagh:2017wot} to uniquely determine a N$^3$LO PN spin-orbit binary Hamiltonian.
We calculated the effective gyro-gravitomagnetic ratios as they would enter the \texttt{SEOBNR}~\cite{Bohe:2016gbl,Babak:2016tgq,Cotesta:2018fcv,Ossokine:2020kjp} and \texttt{TEOBResumS}~\cite{Nagar:2018plt,Nagar:2018zoe} waveform models, and we obtained the gauge-invariant scattering angle and circular-orbit binding energy for aligned spins.
Since the spin-orbit interaction is universal, our results are applicable to generic spinning binaries, e.g., binaries containing neutron stars.

In Fig.~\ref{fig:bindingenergy} we compared the EOB-resummed binding energy against NR results.
The EOB resummation shows a nice convergent behavior towards NR (for aligned spins) even in the strong field regime, which is usually not expected for asymptotic series expansions like the PN one.
More importantly, the new contribution obtained in this {\paper} roughly halves the gap to NR in the high-frequency regime compared to earlier N$^2$LO results for $q=1$.
This indicates that improved (resummed) analytical predictions based on our result can be trusted to higher frequencies, which may alleviate the need for longer and computationally very expensive NR waveforms.
Hence, it is of particular value and urgency to improve the accuracy of the PN-approximate analytic part of GW models.

A clear avenue for future work is to consider higher orders in spin (and higher multipoles).
In particular, in a forthcoming publication, we fix the $S_1S_2$ couplings at N$^3$LO (5PN order) for aligned spins using known self-force results.
It seems reasonable to expect that complete quadratic-in-spin contributions at N$^3$LO, for BHs, for aligned and perhaps even generic spins, should be within reach of first-order self-force computations.  These would require both further self-force observables and new conceptual developments, in particular, generalizations of first-law relations to include higher orders in spin and higher multipoles, and to the case of generic spin orientations (the precessing case).
Future first-order self-force results for unbound orbits may also enable obtaining spin effects to fourth order in the weak-field (post-Minkowskian) approximation---for generic masses and velocities---for BH scattering events (only the second order is currently known~\cite{Bini:2018ywr}).
While this scenario is unlikely to be of astrophysical relevance, it is still very interesting to consider from a conceptual point of view: after all, scattering encounters are the most elementary form of interaction.

\section{Acknowledgments}

We are grateful to Maarten van de Meent for helpful discussions, and to Alessandra Buonanno for comments on an earlier version of this manuscript.
We also thank Sergei Ossokine and Tim Dietrich for providing NR data for the binding energy and for related useful suggestions.


\input{4p5_refs}

\end{document}

%% file: 4p5_refs.tex
%